\documentstyle[sprocl]{article}

\input{psfig}

\def\be{\begin{equation}}
\def\ee{\end{equation}}
\def\bea{\begin{eqnarray}}
\def\eea{\end{eqnarray}}

\begin{document}

 \title{RECENT DEVELOPMENTS IN SUPERSYMMETRIC DARK MATTER}

\author{ACHILLE CORSETTI}

\address{Department of Physics, Northeastern University\\
Boston, MA 02115, USA
\\E-mail: corsetti@neu.edu}

\author{PRAN NATH}

\address{Department of Physics, Northeastern University,
Boston, MA 02115, USA\footnote{: Permanent address}\\
 Physikalisches Institut, Universitat Bonn,
Nussallee 12, D-53115 Bonn, Germany\\
Max-Planck-Institute fuer Kernphysik, Saupfercheckweg 1,
D-69117 Heidelberg, Germany\\
E-mail: nath@neu.edu}


\maketitle\abstracts{A brief review is given of some of the recent
developments in the theoretical analyses of supersymmetric dark matter.
These include the effects of uncertainties in the
wimp velocity and wimp density and of the effects of uncertainties in
the quark densities of the proton. Also analyzed are the effects of
non-universalities in
 the gaugino sector and their effects on determining
 the nature of cold dark matter, i.e., if the neutralino
 is bino like, higgsino like, or wino like.
 The maximum and the minimum elastic neutralino proton cross sections
 are discussed and a comparison of the direct and the indirect
 detection arising from the capture and annihilation of neutralinos
 in the core of the earth and the sun is given. Some of the other recent
 developments are summarized.}

\section{Introduction}
 Because of the recent significant activity in dark matter searches
 on the experimental side\cite{dama,cdms,hdms}
 there is renewed interest in the theoretical analyses
 of dark matter which are significantly more refined than
 in the previous years.  Among the refinements is the inclusions
 of the effects
 of uncertainties in the input parameters in the theoretical
 predictions of event rates and of the neutralino proton cross sections
 as well inclusion of the effects of non-universalities, CP violating
 effects and the effects of coannihilation.
The content of the paper is as follows:
In Sec.2 we discuss the effects of uncertainties in the analyses
of dark matter. These include the effects of uncertainties in the
wimp velocity and in the wimp relic density, and the effects of
uncertainties in the quark densities in the proton in the
analyses of dark matter. In Sec.3 we give a discussion of the
maximum and the minimum elastic neutralino-proton cross-sections.
In Sec.4 we discuss the effects of non-universalities and
specifically the non-universalities in the gaugino sector on
dark matter analyses. A brief discussion of the effects of
 $\mu$ on the composition of the neutralino and its
role in determining the nature of cold dark matter, i.e., if
it is dominantly a bino, a wino or a higgsino is given in Sec.5.
A comparison of the direct and the indirect detection of dark matter
is given in Sec.6. In Sec.7 we give a discussion of the annual
modulation effect in the direct detection of dark matter.
 In Sec.8 we give a brief discussion of the
effects of CP phases on dark matter. Conclusions are given in
Sec.9.
\begin{figure}[t]
\psfig{figure=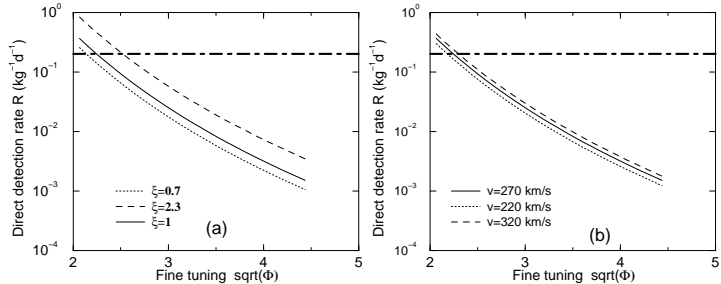,height=1.5in}
\caption{An exhibition of the effect of the variation of the wimp
velocity and wimp relic density on the event rates in a
Ge detector as a function of the fine tuning parameter $\Phi$.
Taken from Ref.$^9$ }
\end{figure}
\section{Uncertainties in Theoretical Analyses of Dark Matter}
The direct detection of dark matter has been investigated by
many authors ( see Ref.\cite{direct} for some of the
recent works).
 Several types of uncertainties enter in these analyses
 such as the effects of variations in the local wimp density, the effects of
variations in the wimp velocity range, and the effects of uncertainties in
the quark densities in the nucleons. Other effects not considered here
are the uncertainties in the nuclear form factors, and
effects of halo models\cite{sikvie} on the event rates in dark matter detection.
We begin with a discussion of the analyses of uncertainties in wimp
density\cite{an1} and velocity\cite{roszkowski,bottino1,corsetti1}.
The current range of
local wimp density lies in the range\cite{gates} $(0.2-0.7) GeV cm^{-3}$.
Defining   $\xi=\rho_{\tilde\chi_1^0}/\rho_0$  one can parameterize
the local wimp density in term of $\xi$ which for
$\rho_0=0.3 GeVcm^{-3}$ gives  $0.7\leq \xi\leq 2.3$.
For the wimp velocity one typically assumes a Maxwellian velocity
distribution for the wimps. Estimates of the rms wimp velocity
range give\cite{knapp}
$ v=270\pm 50$ km/s.
The analysis is carried out  within the framework of the minimal
supergravity model (SUGRA)\cite{chams}.
 The soft SUSY breaking sector in the minimal
version mSUGRA of this theory
 is given by $m_0,m_{1/2}, A_0$, and $\tan\beta$ where, $m_0$ is the
 universal scalar mass, $m_{1/2}$ is the universal gaugino mass,
 $A_0$ is the universal trilinear coupling, and $\tan\beta=<H_2>/<H_1>$
 where $H_2$ gives mass to the up quarks and $H_1$ gives mass to
 the down quarks and the leptons. The Higgs mixing parameter
 $\mu$ is determined by the constraint of radiative breaking of
 the electro-weak symmetry. It is also useful to define a fine
 tuning parameter $\Phi$ so that\cite{ccn}
 $\Phi=(\mu^2/M_Z^2+1/4)^{1/2}$.
 The fine tuning parameter defines how heavy the SUSY spectrum gets.
The effects of the uncertainly in the
event rate as a function of the fine tuning parameter
due to variations in the
wimp relic density and in the wimp velocity are shown in
Fig.1\cite{corsetti1}.
One finds that the effects of this type of uncertainty can lead
to a variation in the rates by a factor of 2-3. \\

Next we discuss the uncertainties in neutralino-proton cross-section
arising from errors in the quark densities in the
proton\cite{bottino2,efo,corsetti2}.
The basic interaction governing the $\chi -p$ scattering with
CP conservation is the four Fermi interaction given by
${\cal L}_{eff}=\bar{\chi}\gamma_{\mu} \gamma_5 \chi \bar{q}
\gamma^{\mu} (A P_L +B P_R) q
+ C\bar{\chi}\chi  m_q \bar{q} q
+D  \bar{\chi}\gamma_5\chi  m_q \bar{q}\gamma_5 q$.
For heavy target materials the neutralino-nucleus scattering is dominated
by the scalar interactions which is controlled by the
scalar $\chi -p$  cross-section where
\begin{equation}
 \sigma_{\chi p}(scalar)=\frac{4\mu_r^2}{\pi}
 (\sum_{i=u,d,s}f_i^pC_i+\frac{2}{27}(1-\sum_{i=u,d,c}f_i^p)
 \sum_{a=c,b,t}C_a)^2
 \end{equation}
 Here $\mu_r$ is the reduced mass and  $f_i^p$ (i=u,d,s quarks)
 are the (u,d,s) quark densities defined by
 $m_pf_i^p=<p|m_{qi}\bar q_iq_i|p>$ (i=u,d,s).
In the above C is the strength of the scalar interaction and
consists of s channel contributions from the higgs $h^0, H^0$
exchange and t channel contributions from the sfermion exchange
so that $C=C_{h^0}+C_{H^0}+C_{\tilde{f}}$. The uncertainly
in the theoretical predictions of $\sigma_{\chi p}(scalar)$
 is dominated by the uncertainty in the quark densities $f_i^p$.
To study the uncertainties in $f_i^p$ it is best first to solve
 the quark densities analytically in terms of some judiciously
chosen parameters. One finds\cite{corsetti2,corsetti3}

\begin{eqnarray}
f_{(u,d)}^p=\frac{m_{(u,d)}}{m_u+m_d}(1\pm\xi)\frac{\sigma_{\pi N}}{m_p},~~~
f_s^p=\frac{m_s}{m_u+m_d}(1-x)\frac{\sigma_{\pi N}}{m_p}
\end{eqnarray}
where the parameters $\xi$, x and $\sigma_{\pi N}$ are defined by

\begin{eqnarray}
 \xi=
 \frac{<p|\bar uu-\bar dd|p>}{<p|\bar uu+\bar  dd|p>},~~~
x= \frac{<p|\bar uu+\bar dd-2\bar ss|p>}{<p|\bar uu+\bar  dd|p>}\nonumber\\
 <p|2^{-1}(m_u+m_d)(\bar uu+\bar dd|p>=\sigma_{\pi N}
\end{eqnarray}
Similarly one can analytically solve for the quark densities in the
neutron and the analytic relations provide an interesting connection
between the quark densities in the proton and in the neutron.
One finds that independent of the details of any input
one has\cite{corsetti2}$f_u^pf_d^p=f_u^nf_d^n$.
One can use the analysis on the baryon mass splittings\cite{hcheng}
to determine the ratio $\xi/x$. One finds\cite{corsetti2} $\xi/x=0.196$.
Using the determination of x from lattice gauge
analyses\cite{borasoy,dong,fukugita}
one finds $\xi=0.132\pm 0.035$. Additional uncertainties can
arise from the quark mass ratios. Here results from
chiral perturbation theory\cite{gl,saino} give
$\frac{m_u}{m_d}=0.553\pm 0.043$, $\frac{m_s}{m_d}=18.9\pm 0.8$.
Using the inputs above one finds
$f_u^p=0.021\pm 0.004$,
$f_d^p=0.029\pm 0.006$,
$f_s^p=0.21\pm 0.12$ and
$f_u^n=0.016\pm 0.003$,
$f_d^n=0.037\pm 0.007$,
$f_s^n=0.21\pm 0.12$.
\begin{figure}[t]
\psfig{figure=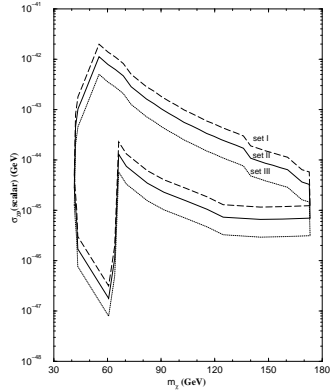,height=1.5in}
\caption{The effect on $\sigma_{\chi -p}$ of the variations in
the quark densities around the central value given in Sec.2}
\end{figure}
\section{Maximum and Minimum Neutralino-proton Cross Sections}
We discuss now the numerical results of the neutralino proton
cross-sections with the quark densities as discussed in Sec. 2.
In Fig.2 $\sigma_{\chi -p}$ is plotted exhibiting the effects
of the variations of the quark density as a function of $m_{\chi}$.
The above analysis shows that the $\chi -p$ cross section cannot
be computed to an accuracy of better than a factor of 3-5 with
the current level of uncertainties in the input data.
Fig.2 also gives a plot of the maximum and the minimum of
$\sigma_{\chi -p}$ as a function of the neutralino mass.
 In the analysis of Fig.2 we have allowed $m_0$ and
$m_{1/2}$ to vary in the range up to 1 TeV and the constraint
from the flavor changing neutral current process
$b\rightarrow s\gamma$ is imposed\cite{fcnc}. The analysis of this
figure does not include the effects of the spin dependent
contribution which could change the scatter plot. Specifically, the
minimum cross sections are sensitive to the inclusion of the
spin dependent part. Similar analyses
can be found in other recent references\cite{mandic,lahanas}.\\

\section{Effects of Non-universalities on Dark Matter}
mSUGRA is based on the assumption of a flat Kahler potential.
However, the nature of physics at the Planck scale is not
fully understood. Thus in general one should allow for the
possibility of a curved Kahler potential. This would lead to
non-universalities in the scalar sector. However, there are
stringent constraints on the types of non-universalities allowed
in the scalar sector due to the FCNC constraint. For example,
 the FCNC constraint in very strong on
the amount of non-universality allowed in the first vs the
second generation sector. However, this constraint is not so
strong for the Higgs sector and for the third generation sector.
Effects of these constraints have been studied in detail
in Refs.\cite{nonuni,accomando}. One finds that in general the presence of
non-universalities can increase the cross-sections by a
factor of 10 or more.

In addition to modifying the Kahler potential
Planck scale physics can also modify the gauge kinetic
energy function. In general the gauge kinetic energy function
transforms as the symmetric product of two adjoint representations.
For the case of $SU(5)$ one finds that the gauge kinetic energy
function $f_{\alpha\beta}$ transforms as the
symmetric product of ${\bf 24\times 24}$ in SU(5) which
contains the following irreducible representations of
$SU(5)$:
\begin{equation}
({\bf 24}\times {\bf 24})_{symm}={\bf 1}+{\bf 24}+{\bf 75}
+{\bf 200}
\end{equation}
The SU(5) singlet in the product on the right hand side of Eq.(4)
leads to a universal gaugino mass while, the additional terms
generate non-universalities in the gaugino masses at the
GUT scale. Thus the $SU(3)_C\times SU(2)_L\times U(1)$
gaugino masses at the GUT scale are in general admixtures of
all the allowed representations. This admixture leads to
the following relation for the gaugino masses at the GUT scale

\begin{equation}
\tilde m_i(0)=m_{\frac{1}{2}}(1+ \sum_r c_r n_i^r)
\end{equation}
where $n_i^r$ depend on one of the representations on the right
hand side in Eq.(4) and on the subgroup i.
In addition to the appearance of gaugino mass non-universalities
of the above type in supergravity models, one also finds quite
naturally non-universalities in a broad class of string models:
heterotic, Horava-Witten and as well in brane models
based on TypeI/Type IIB string compactifications.
 Now the value of $\mu$ is
in general very sensitively dependent on non-universalities.
For the case of non-universalities in the scalar sector
one can exhibit in an analytic fashion the dependence of $\mu$ on
non-universalities in the Higgs sector and in the third generation
sector. The analysis shows that $\mu$ is
sensitively dependent on the non-universalities and their effects
can significantly decrease the value of
$\mu$ and thus affect gaugino vs higgsino composition of the
neutralino. An analysis of the effects of non-universalities in
the scalar sector is given in Ref.\cite{nonuni}.
A similar phenomenon occurs for the case of gaugino sector
non-universalities. Here also one can exhibit the dependence
of $\mu$ on non-universalities. One finds\cite{corsetti2}

\begin{equation}
  \tilde \mu^2= \mu^2_{0}+\sum_r \frac{\partial\tilde\mu^2}
  {\partial c_r} c_r+ O(c_r^2)
\end{equation}
where $\partial \mu_{24}^2/\partial c_{24}>0,
\partial \mu_{75}^2/\partial c_{75}>0,
\partial \mu_{200}^2/\partial c_{200}<0$.
Again with an appropriate choice of the sign of $c_r$ one
finds that $\mu$ becomes smaller relative to its universal value.
An analysis of the effects of non-universalities in the gaugino
sector on $\sigma_{\chi -p}$ is given in Ref.\cite{corsetti2}.
Fig.3 gives a typical illustration of the effects of gaugino
non-universality showing that a significant enhancement of
$\sigma_{\chi -p}$ can occur in the presence of non-universalities.
\begin{figure}[t]
\psfig{figure=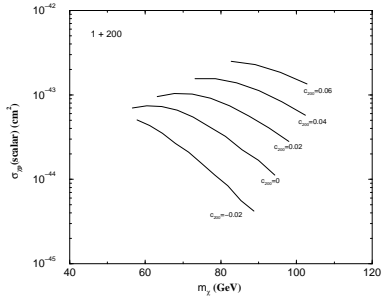,height=1.5in}
\caption{Exhibition of the effects of non-universalities on
$\sigma_{\chi -p}$ as a function of the
neutralino mass for different values of $c_{200}$.}
\end{figure}

\section{Bino, Higgsino, and Wino Dark Matter}
The neutralino is in general a mixture of gauginos and higgsinos,
\begin{equation}
\chi_1=X_{11} \tilde B+ X_{12}\tilde W_3 + X_{13}\tilde H_1
+ X_{14} \tilde H_2
\end{equation}
where  $\tilde B$ is the bino, $\tilde W$ the neutral wino,
and $\tilde H_1, \tilde H_2$ are higgsinos.
Now an analysis of the neutralino mass matrix shows that for large
values of the Higgs mixing parameter $\mu$ one finds that
the neutralino is essentially a bino and this situation is realized
over a large part of the SUGRA parameter space\cite{bino}. Thus SUGRA models
in general predict a bino like cold dark matter at least over
a major portion of the parameter space. However, in certain
limited regions of the SUGRA parameter space $\mu$ can become
small. In this case  the higgsino and the wino components can
become large and one may have cold dark matter
which is higgsino like or wino like.
This phenomenon can be easily understood by examining the
$|\mu|>>M_Z$ limit of the components $X_{1n}$.  Here one finds
that the bino component in the neutralino is given by
$ X_{11}\simeq 1-(M_Z^2/2\mu^2)sin^2\theta_W$,
the wino component is of size
$X_{12}\simeq (M_Z^2/2m_{\chi_1}^2\mu)sin2\theta_W sin\beta$,
and the higgsino components have the sizes
 $ X_{13}\simeq -(M_Z/\mu)sin2\theta_W sin\beta$, and
$X_{14}\simeq (M_Z/\mu)sin2\theta_W sin\beta$.
The above illustrates the sensitive dependence of
the bino, the wino and the higgsino components on $\mu$. One
consequence of the effect of large higgsino components is
that the scalar cross section which depends on the
product of the gaugino and the higgsino components  increases as the
higgsino components increase\cite{nonuni}.
An analysis of CDM with large higgsino components in the
context of MSSM is given in Ref.\cite{drees} and with large
higgsino and wino components in the context of anomaly mediated breaking of
supersymmetry in Ref.\cite{wells}

\section{Comparison of Direct and Indirect Detection}
 Indirect detection is complementary to the direct detection in the
 search for dark matter. The most interesting indirect signal
 for dark matter arises from the capture and the subsequent
 annihilation of the neutralinos in the center of the Sun and
 the Earth. Some of the remnants in the annihilation of the
 neutralinos are the neutrinos which propagate and undergo charged
 current interactions in the rock surrounding the detector
 and produce upward moving muons. The outgoing muon flux
 can be written in the form $\Phi_{\mu}=\Gamma_{A}f$ where
 $\Gamma_{A}$ is the $\chi_1-\chi_1$ annihilation rate in
 the center of the Earth or the Sun and f is the product of remaining
 factors. It is $\Gamma_{A}$ which is the quantity sensitive
 to SUSY. One can parameterize $\Gamma_A$ by\cite{greist}
 \begin{equation}
 \Gamma_A=\frac{C}{2} tanh^2(t/\tau)
 \end{equation}
Here C is the neutralino capture rate, t is the lifetime of the Earth or
the Sun, and $\tau =(CC_A)^{-1/2}$ where $C_A$ is determined by the
wimp annihilation cross section. An equilibrium between
capture and annihilation is reached for the case $t >>\tau$
and in this case one has $\Gamma_A\sim \frac{C}{2}$.
The equilibrium condition is strongly dependent on the
susy parameter space. For the case of the sun the equilibrium condition
is satisfied essentially over all of the parameter space while for
the Earth it is satisfied only over part of the parameter space.
This disparity shows up very strongly in the profile of the muon
flux from the Earth and the Sun. In Fig.4 we exhibit the muon flux
for the Earth and the Sun plotted against the direct detection rate for
the germanium detector. The analysis of Fig.4 shows that for
 relatively large direct detection rates, it is the indirect
 detection from the Earth which is competitive with the direct detection
 while for relatively
 low direct detection rates it is the indirect detection from the
 Sun which is competitive with the direct detection. Thus indirect
 detections from the Earth and the Sun are complementary.
 For other recent analyses of indirect detection see
 Refs.\cite{berez,wilczek,edsjo}.
\begin{figure}[t]
\psfig{figure=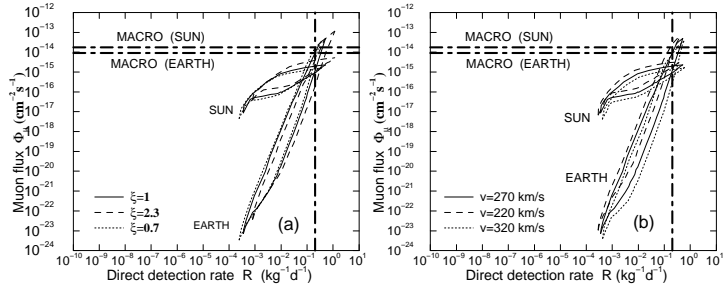,height=1.5in}
\caption{A comparison of the direct and the indirect detection with a plot of muon flux of the Earth and the Sun vs the direct detection rates. Taken from
Ref.$^9$ }
\end{figure}

\section{Annual Modulation Effect in Direct Detection}
The annual modulation effect in the direct detection  of dark matter
is a potentially important signal for the observation of WIMP like
dark matter. The effect arises due to the periodicity of
 the velocity of the Earth $v_E$ relative to the galaxy,i.e.,
 $v_E=v_S+v_0 cos\gamma cos\omega (t-t_0)$ where $v_0(=30km/s)$ is
Earth's orbital velocity around the Sun, $v_S (=232km/s)$ is the
Sun's velocity relative to the galaxy, $\gamma (\simeq 60^0)$ is
the inclination of Earth's orbit relative to the galactic plane,
and $t_0$ is June 2. The effects of this motion can produce a
modulation effect of about $7\%$ in the scattering event rates
for wimps. The DAMA experiment claims to see an effect\cite{dama}.
The annual modulation signal has been analyzed theoretically in
several papers\cite{bottino4,an1,green,vergados}.
In the future one expects further data on this exciting possibility.
Further, one needs to reduce the current ambiguities in the
theoretical predictions of dark matter such  as those discussed
in Sec.2. This requires more accurate chiral perturbation and
lattice gauge analyses.

\section{Effects of Large CP Phases on Dark Matter}
It is well known that supersymmetry brings in new sources of CP
violation since the soft SUSY parameters are in general
complex. These CP phases induce additional corrections to
the electric dipole moments of the electron and of the neutron.
We consider here the case where the CP phases are large and the
edm constraints are satisfied.
(For an abbreviated set of references on large phases
see Ref.\cite{largephases}). Large  phases affect
the analysis of dark matter and the
detailed  analyses of event rates for direct detection
show that the effects of the phases can change the event
rates by an order of magnitude or more\cite{fos,cin}.
However, with the inclusion of the
emd constraints one finds that the effects are significantly
reduced although they are still substantial\cite{cin}.
Large phases also induce a mixing between the CP even and the CP odd
higgs states\cite{pilaftsis,in2} which can affect dark matter\cite{in2}.

\section {Conclusion}
In this review we have presented the very recent
developments in the theoretical analyses of supersymmetric dark
matter. These include the effects of uncertainties on the
wimp density and on the wimp velocity for the Milky Way wimps,
 effects of uncertainties in
quark densities in the theoretical predictions of
$\sigma_{\chi- p}$, and the effects of non-universalities
 in the gaugino sector.
 A analysis of the maximum and of the minimum cross sections for
$\chi-p$ scattering was given.
We have also given a comparison of the direct and of the indirect detection
of dark matter and  discussed briefly other recent developments
for the detection of dark matter. An important topic not discussed
is the subject of coannihilation\cite{coanni} which can
significantly extend the domain of the allowed neutralino masses.
In the future one expects that more
sensitive detectors\cite{spooner,genius} will be able to probe more deeply
 the parameter
space of SUGRA and other SUSY models.

\section*{Acknowledgments}
One of us(PN) wishes to thank the Physics Institute at the University of Bonn
and the Max Planck Institute, Heidelberg, for hospitality
and acknowledges support from an Alexander von Humboldt award.
This research was supported in part by NSF grant PHY-9901057.

\end{document}